
\vskip 3cm
\input vanilla.sty
\magnification=1200
\nopagenumbers
\headline={\ifnum\pageno=1\hfil\else\hss\tenrm -- \folio\ --\hss\fi}

\def\wt{\widetilde}

\def\noi{\noindent}

\TagsOnRight
\mathsurround=1pt
\line{Preprint {\bf SB/F/93-214}}
\hrule
\vskip 2.5cm
\centerline{\bf ``TOPOLOGICAL SELECTION OF WORLD MANIFOLDS FOR A}
\centerline{\bf P-BRANE IN A BF--FIELD''}
\vskip 2cm

\centerline{J. P. Lupi, A. Restuccia and J. Stephany}
\vskip 4mm

\centerline{\it Universidad Sim\'on Bol\'{\i}var,Departamento de F\'{\i}sica}
\centerline{\it Apartado postal 89000, Caracas 1080-A, Venezuela}
\centerline{\it e-mail: stephany{\@}usb.ve}

\vskip 1cm

{\narrower\flushpar{{\bf Abstract}

\vskip 3mm

\noi
We study the interaction  between a $p$-brane and $BF$
system  constituted by a $(p+1)$-form and a $n-(p+2)$-form B with a metric
independent action on a manifold $M^n$. We identify the allowed $(p+1)$-world
manifolds sweeped
by the $p$-brane as those restricted by the topological condition of being
the boundaries of chains in $M^n$. We find the general solutions of the
equation of motion for the field configurations. The solutions for the $B$
field are closely related to the external field configurations that describe
the defects needed for the computation of soliton correlation functions.The
particular cases of the particle and string evolution are discussed with  some
detail.}\par}

\vskip 3cm

\hrule
\bigskip
\centerline{\bf UNIVERSIDAD SIMON BOLIVAR}

\newpage
\baselineskip=16pt

\newpage
The study of topological field theories [1] has been intensily developed  in
recent times, mainly in relation with their very interesting mathematical
properties. The possibility of computing non-trivial correlation functions in
these models [1] [2], which are expresed in terms of objects of topological
nature like knot invariants and polynomials, is an important reason for this
emphasis. But also from the physicist's poin of view, topological field
theories have a strong interest. As isolated models, due to their peculiar
properties, they provide an excellent environment for testing some recent
developments in quantum field theory, such as functional methods [3], BRST
quantization [4] [5], stochastic approaches [6], etc.

As elements of interacting models, topological field theories may modify the
behaviour of other fields in the system with definite consequences on the
particles of the spectrum. Given a topological field theory written in terms
of fields $A^b$ interacting with fields $\phi^i$, the restricted partition
function ($a$ and $i$ generic indices)
$$
Z_{top}[\phi_i]=\int\Cal{D}\mu (A^a)e^{-S_{eff}[\phi ,G,A]} \tag 1
$$
does not depend on the background metric $G_{\mu\nu}$. In Eq. (1)
$\Cal{D}\mu (A^a)$ is the adequate measure for the system and
$S_{eff}[\phi ,G,A]$ grows out of the classical action
$$
S[\phi,G,A]=S_{top}[G,A]+S_{int}[\phi,G,A] . \tag 2
$$
Then the complete partition function factorizes in the form
$$\align
Z &=\int\Cal{D}\phi_i\Cal{D}\mu
(A^a)e^{-S_{eff}[\phi,G,A]-S_{matter}[\phi,G]}\\
&=\int\Cal{D}\phi_iZ_{top}(\phi_i)e^{-S_{matter}[\phi ,G]} \tag 3
\endalign
$$
showing  the non-trivial influence of the topological fields on the
matter fields. Moreover, this factorization should even be true for the case
with a dynamical $G_{\mu\nu}$, after circumventing the obstacles presented in
its quantization.

The mechanism discussed above appears to provide the best tool for an
adequate realization of 3-dimensional anyons [7]. For the first quantized
formulation of the interaction of two or more anyons, a ``statistical'' field
with a Chern-Simons  (topological) action has been used in order to generate
in the functional  integral the weight factors corresponding to the different
representations  of the fundamental group of the configuration space
( $ \text{\bf R}^3 \times \text{\bf RP}^1$ for two anyons).

{}From a slightly shifted point of view Polyakov [8] showed explicitly that
the  propagator for bare bosonic particles interacting
with a pure Chern-Simons field is fermionic. This result was generalized  for
strings interacting with a $BF$ [9] field in 4 dimensions in Ref. [10].

In this letter we will pursue further in this direction the analysis of
topological field theories by considering the system composed by an abelian
topological $BF$ model interacting with the sources of ``first quantized''
matter which couple consistently to it. As a first but important step we will
concentrate in the classical aspects of this system and we will find the
general solutions to the dynamics of the systems. In particular, we will show
that the interaction with the $BF$ fields imposes topological restrictions on
the otherwise arbitrary $(p+1)$-geodesics sweeped by the $p$-branes in their
evolution.

The Abelian $BF$ system [9] on an $n$-dimensional manifold $M^n$ is described
by the metric independent action
$$
S_{BF}(n,p)=\int_{M^n}B_{n-p-2}\land dA_{p+1} \tag 4
$$
with $A$ and $B$ forms, the subscript indicating their rank. The partition
function corresponding to this system is topological (i.e. metric
independent) [5] [9] [11].

This system couples naturally with a $p$-brane which sweeps a $(p+1$-world
manifold on $M^n$ through its evolution. The $p$-brane action and the
interaction terms are
$$\align
S_{p-brane}&=\int d^{p+1}\sigma\sqrt{det|g_{ij}|}\tag 5\\
S_{int} &=-e\int_{W^{p+1}}A_{p+1}.\tag 6
\endalign
$$
Here $g_{ij}=G_{\mu\nu}\partial_ix^\mu \partial_jx^\nu$, with $G_{\mu\nu}$
the metric on $M^n$, is the induced metric on the world manifold $W^{p+1}$.

The complete action we will consider is
$$
S=S_{BF}(n,p)+S_{p-brane}+S_{int}\tag 7
$$
The field equations are obtained by annihilating the variations of $S$ with
respect to:
\item{a)}$B_{n-p-2}$, implying that $A_{p+1}$ is a flat connection.
$$
F_{p+2}=dA_{p+1}=0 \tag 8
$$
\item{b)}$x^\mu$, yielding, using a), the trajectories of extremal world
volume (i.e. geodesics).
\item{c)}$A$, imposing the topological restriction
$$
\int_{M^n} B_{n-p-2}\land d\varphi_{p+1}-e
\int_{W^{p+1}}\varphi_{p+1}=0 \tag 9
$$
when applied to an arbitrary test ($p+1$)-form $\varphi$ of compact support.
Here we are assuming that the adequate boundary restrictions, if any, are
satisfied by the fields.

Before discussing the solution of these equations in all generality, let
us discuss some illustrating particular cases.

In a previous letter [12], the problem of the interaction of a massive test
particle with $BF$ fields in a 2-dimensional manifold $M^2$ was considered.
In this case, the action (7) for the whole system written in components
notation is
$$\align
S=&-m\int_{W^1}d\tau \sqrt{\dot{x}^2}+e\int_{M^2}d^2\xi\int d\tau A_\mu (\xi )
\dot{x}^\mu(\tau )\delta^2(\xi -x(\tau ))\\
&+\epsilon^{\mu\nu}\int_{M^2 }d^2\xi B\partial_\mu A_\nu  \tag 10
\endalign
$$
Here $\xi^\mu$ are the local coordinates of the manifold $M^2$ and $W^1$ is
the world line of the particle, described locally by $x^\mu (\tau)$. The
equation of motion obtained performing variations with respect to $B$ is
again (8), or in component notation, $\epsilon^{\mu\nu}\partial_\nu A_\nu =0$,
stating that the connection $A$ is flat. Thus, there are classicaly no forces
on the particle. Variations with respect to $x^\nu$ tell us that the world
line $W^1$ is a geodesic on $M^2$.

The interesting aspects of the problem appear when we solve the third
equation of motion. Variations with respect to $A$ give:
$$
e\int d\tau \dot{x}^\mu(\tau )\delta (x(\tau ) -\xi ))=
\epsilon^{\mu\nu}\partial_\nu B \tag 11
$$
This equation is metric independent. As we show below [12], it restricts the
set of allowed geodesics $W^1$ of the particle to those curves which are
homologous to zero (i.e. boundaries of 2-chains). To see this, we apply to
both sides of (11) a test 1-form $\varphi =\varphi_\mu (\xi )d\xi^\mu$ of
compact support. We obtain the equality
$$
e\int_{W^1}\varphi = \int_{M^2}Bd\varphi \tag 12
$$
which is a particular case of (9).

This equation can only be solved [12] by taking $W^1$ homologous to zero on
$M^2$. Under these conditions, it can be proved [12] from the mod 2
intersection invariant theory of $p$ and 1-chains on a $(p+1)$-dimensional
manifold that $W^1$ divides $M$ in two manifolds, $M_+^2(M_-^2)$ at right
(left) of their oriented boundary $W^1$. We obtain an ``elementary solution''
to equation (12) by taking the field $B$ to be a step solution that is of
constant value $B_+(B_-)$ on $M_+^2(M_-^2)$, with $B_+-B_-=1$. Using the
Stokes theorem we inmediatly get
$$
\int_{M^2}Bd\varphi =\int_{M^2_+}B_+d\varphi +\int_{M_-^2}B_-d\varphi =
(B_+-B_-)\int_{\partial M^2_+}\varphi =\int_{W^1}\varphi \tag 13
$$
In Ref. [12] it was also shown how to construct globally a step function $B$
on  any oriented manifold when it is divided by a composition of several
curves  $W_i^1$ homologous to zero.

As a second example let us consider the problem of a particle in
a 3-dimensional manifold $M^3$. The action (7) reads in this case with
analogous  notation $$\align S=&-m\int_{W^1}d\tau
\sqrt{\dot{x}^2}+e\int_{M^3}d^3\xi\int_{W^1}d\tau A_\rho  (\xi )\dot{x}^\rho
(\tau )\delta^3(\xi -x(\tau ))\\ &+\epsilon^{\mu\nu\rho}\int_{M^3}d^3\xi
B_\mu\partial_\nu A_\rho .\tag 14 \endalign $$
Again, the connection must be flat and the trajectories geodesics. The
topological restrictions on the geodesics are determined by the equation
obtained by performing variations with respect to $A$ (and are a particular
case of (9))
$$
e\int_{W^1}d\tau \dot{x}^\rho \delta^3(\xi -x(\tau ))=
\epsilon^{\mu\nu\rho}\partial_\nu B_\mu \tag 15
$$
Since this equation is linear, the general solution for $B$ will be a
superposition of solutions to the homogeneous equation and a particular
solution. The solution to the homogenous equation $dB=0$ is a closed
form. This fact can be restated by saying that (15) is invariant under a
transformation of the form
$$
B+ = B + {\text{ closed form}} \tag 15a
$$
If the closed form is also exact we may consider (15a) as a gauge
transformation. If not, the two sides of equation (15a) belong to
different  cohomology clases  in the space of solutions.

 For the sake of clarity let us
restrict further our example and consider the  case when $M^3=\text{\bf R}^3$
and $W^1$ lies entirely in the $(\xi^0,\xi^1)$  plane. We call this plane
$S^2_1$ for later generalization (see Fig. 1). We  observe that $S^2_1$
divides $M^3$ into two regions $M_{1+}^3$ (the upper  semi-space) and
$M_{1-}^3$.

Under these suppositions and taking the  particular solution to satisfy
$B_0^{part}=B_1^{part}=0$, the topological equation (15) reads:
$$\align
-e\delta (\xi^2)\int_{W^1}d\tau x^\alpha (\tau )\delta^2(\xi -x(\tau ))&=
\epsilon^{\alpha\beta}\partial_\beta B_2^{part} \tag 16\\
&\ \ \ \ (\alpha ,\beta )=0,1
\endalign
$$
Except for the factor $\delta (\xi^2)$, this is the equation (11) of our
previous discussion. The elementary solution to this equation can be
constructed [12] only for $W^1$ homologous to zero on $S^2_1$. In this case,
$W^1$ is the boundary of a 2-chain and divides the $(\xi^0,\xi^1)$ plane
$S_1^2$ in two regions: $S_{1+}^2(S_{1-}^2)$ at the right (left) of the
curve $W^1$.

The only non vanishuing component of the particular solution $B^{part}$ is
$$
B_2^{part}=e\theta_{(S^2_{1+})}\delta (\xi^2) \tag 17
$$
where $\theta_{(S^2_{1+})}$ is step function defined (when
$W^1=\partial S^2_{1+}$) as
$$
\theta_{(S^2_{1+})}=\cases
1 &\text{on $S^2_{1+}$}\\
0 &\text{on $S^2_{1-}$}
\endcases
$$

We now introduce a second 2-dimensional submanifold $S^2_2$ such that is the
boundary of  a 3-chain in $M^3$, $S_2$ intersects $S_1$ transversally and
$W^1=S_1^2\cap S_2^2$ (see Fig. 2).Two $(n-1)$-submanifolds of $M^n$ are said
to intersect transversally if at  each point $x$ of their intersection the
tangent space of each one generates  the whole of the tangent space of $M^n$.

The solution (17) may then be rewritten in the form
$$
B_1^{part}=e\Theta_{S^2_1}d\Theta_{S^2_2} \tag 17a
$$
where $\Theta_{S_1^2}$ is the step function
$$
\Theta_{S^2_1}=\cases
1 &\text{on $M^3_{1+}$}\\
0 &\text{on $M^3_{1-}$}
\endcases
\tag 18
$$
(note that $\theta_{(S^2_{1+})}$ is defined only in the 2-dimensional
manifold $S^2_1$, while $\Theta_{S^2_1}$ is defined in the whole space).
Observing that $W^1=S^2_2\cap S_1^2=S^2_2\cap \partial M^3_{1+}=
\partial (M_{2+}^3\cap M^3_{1+})$ we show that (17a) satisfies (9) (or (15)):
$$\align
\int_{M^3}B_1^{part}\land d\varphi_1&=e\int_{M^3}\Theta_{S^2_1}d\Theta_{S^1_2}
\land d\varphi_1=e\int_{M^3_{1+}}d\Theta_{S^1_2}\land d\varphi_1=
e\int_{\partial M^3_{1+}}\Theta_{S^1_2}d\varphi_1\\
&=e\int_{M^3_{2+}\cap \partial M^3_{1+}}d\varphi_1=
e\int_{\partial (M^3_{2+}\cap \partial M^3_{1+})}\varphi_1=
e\int_{W^1}\varphi_1 \tag 19
\endalign
$$
At this point we can return to the more general case when $W^1$ is an
arbitrary geodesic in a general 3-manifold  $M^3$. From the considerations
above it is clear that,  if we take two surfaces $S_i^2$, intersecting
transversally, each one dividing  $M^3$ in two regions $M^3_{i+}$ and
$M^3_{i-}$ and such that  $W^{1}=S_1^2\cap S_2^2$, then equation (17a) still
defines a solution to  our problem.

It is important to observe that our choice of the surfaces $S_i^2$ used to
describe our solution is not uniquely determined by the curve $W^1$. We
may have equally taken other two manifolds $\wt{S}_i^2$ such that: first, both
 divide $M^3$ in two regions $\wt{M}_{i+}^3$ and $\wt{M}_{i-}^3$; second,
$W^1$ is homologous to zero in each one; and third,
$W^1=\wt{S}^2_1\cap \wt{S}^2_2$. It is clear that it is possible to
construct another elementary solution of the form (17a) in terms of the new
objects. Two solutions $B^{part}$ and $\wt{B}^{part}$ obtained from different
choices of the intersecting surfaces have to be related by a transformation
of type (15a). Let us consider the case
$$\align
B^{part}_1 &=\Theta_{S^2_1}d\Theta_{S^2_2}\\
\wt{B}^{part}_1 &=\Theta_{\wt{S}^2_1}d\Theta_{\wt{S}^2_2} \tag 20a
\endalign
$$
with $W^1=S^2_2\cap S^2_1=\wt{S}^2_2\cap \wt{S}^2_1$. Since the support of
the distribution $d\Theta_{{S}^2_i}$ is $S^2_i$ we have
$$\align
B^{part}_1
&=\Theta_{S^2_1}d\Theta_{S^2_2}=\Theta_{\wt{S}^2_1}d\Theta_{{S}^2_2}=d(\Theta_{\wt{S}^2_1}\Theta_{{S}^2_2})-d\Theta_{\wt{S}^2_1}\Theta_{{S}^2_2}=d(\Theta_{\wt{S}^2_1}\Theta_{{S}^2_2})-d\Theta_{\wt{S}^2_1}\Theta_{\wt{S}^2_2}\\

&=d(\Theta_{\wt{S}^2_1}\Theta_{{S}^2_2}-\Theta_{\wt{S}^2_1}\Theta_{\wt{S}^2_2})+\Theta_{\wt{S}^2_1}d\Theta_{\wt{S}^2_2}=\wt{B}^{part}_1 + d\Lambda .\tag 20b
\endalign
$$

The relation between $B^{part}_1$ and $\wt{B}^{part}_1$ is of the form
(15a).The distribution $\Lambda$ that appears above has a geometrical
interpretation: it is the step function with support in the subset of
$\wt{M}^3_{1+}$ enclosed between $S^2_2$ and $\wt{S}^2_2$. It is interesting
to observe that interchanging $S^2_1$ and $S^2_2$ in (17a) one gets another
solution to our field equations corresponding to the opposite orientation of
$W^1$.

We now
return to the general problem of solving equation (9) for the case  of a
$p$-brane in an n-dimensional manifold. In the light of our previous  examples,
we expect that (9) imposes topological restrictions on the class  of allowed
geodesics. Since the structure of equation (9) in any dimension  is the same
as that of equations (11) and (15), its solution is again linked  to the
possibility of constructing a step function with non-vanishing value  over a
submanifold whose boundary is the  world manifold $W^{p+1}$.

In what follows we first obtain an ``elementary solution'' to eq. (9) which
generalizes (17a), then the most general solution is constructed from the
composition of the elementary ones.

In order to construct the ``elementary solution'' mentioned above, we
introduce $n-(p+1)$ submanifolds $S_i^{n-1}$. The $S_i^{n-1}$ are
$(n-1)$-dimensional oriented submanifolds satisfying the following properties:
\item{i)}$S^{n-1}_i$ is the boundary of an $n$-chain \item{ii)}$S^{n-1}_i$
intersects $S^{n-1}_j$ transversally for any $i$ and $j$.

\noi
The world manifold $W^{p+1}$, in addition to being a geodesic, has to be
obtained by the intersection of such a set of $n-(p+1)$ submanifolds
$S_i^{n-1}$.

Since we chose the $S_i^{n-1}$ to be the boundary of a $n$-chain in $M^n$, it
can be shown using the intersection invariant theory that each one of them
divides $M^n$ in two regions, $M^n_{i+}$ and $M^n_{i-}$ (see [12] for the two
dimensional case). We take $M^n_{i+}$  to satisfy $\partial
M^n_{i+}=S_i^{n-1}$ with the orientation of  $\partial M^n_{i+}$ induced by
the orientation of $M_{i+}^n$ in the usual  way consistent with the Stokes
theorem $$
\int_{M^n_{i+}}d\varphi =\int_{S^{n-1}_i}\varphi \tag 21
$$

The elementary world manifold $W^{p+1}$ we will consider is given by
$$\align
W^{p+1}=\partial (M^n_{n-p-1+}&\cap \partial (M^n_{n-p-2+}\cap \cdots \\
&\cap \partial (M^n_{2+}\cap \partial M^n_{1+}))\cdots ) \tag 22
\endalign
$$
and a particular solution of (9) for the $B$ field is then given by following
current
$$
B^{part}_{n-p-2}=e\Theta_{S_1^{n-1}}\delta_{S^{n-1}_2}\land \cdots \land
\delta_{S^{n-1}_{n-p-1}} \tag 23
$$
with $\Theta_{S^{n-1}_i}$ and $\delta_{S^{n-1}_i}$ described as follows:
$\Theta_{S^{n-1}_i}$ is the step function with value $+1$ (0) on
$M^n_{i+}(M^n_{i-})$. $\delta_{S_i^{n-1}}$ is the exterior derivative of
$\Theta_{S^{n-1}_i}$: applying to a test ($n-1$)-form of compact support
$\varphi_{n-1}$ we have
$$\align
\int_{M^n}\delta_{S^{n-1}_i}\land \varphi_{n-1} &\equiv
\int_{M^n}d\Theta_{S_i^{n-1}}\land \varphi_{n-1}=
\int_{M^n}d(\Theta_{S_i^{n-1}}\land
\varphi_{n-1})-\int_{M^n}\Theta_{S_i^{n-1}}d\varphi_{n-1}\\
&=\int_{\partial M^n}\Theta_{S_i^{n-1}}\land
\varphi_{n-1}-\int_{M^n_{i+}}d\varphi_{n-1}=-\int_{S^{n-1}_i}\varphi_{n-1} \tag
24
\endalign
$$
where we use Stokes theorem and the fact that $S^{n-1}_i = \partial
M^n_{i+}-\partial M^n \cap
M^n_{i+}$ This justifies the name of $\delta_{S_i^{n-1}}$. We may then prove
by  recurrence that (23) is a solution of (9) observing that for any test form
$\varphi_{p+1}$
$$\align
\int_{M^n}B_{n-p-2}^{part}\land d\varphi_{p+1} &=
e\int_{M_{1+}^n}\delta_{S_2^{n-1}}\land \cdots \land \delta_{S^{n-1}_{n-p-1}}
\land d\varphi_{p+1}\\
&=e\int_{\partial M^n_{1+}}\Theta_{S_2}\delta_{S_3}\land \cdots \land
\delta_{S^{n-1}_{n-p-1}}\land d\varphi_{p+1}\\
&=e\int_{\partial (M^n_{2+}\cap \partial M^n_{1+})}\delta_{S_3}\land \cdots
\land \delta_{S^{n-1}_{n-p-1}}\land d\varphi_{p+1} \tag 25
\endalign
$$
After $(n-p-1)$ steps we reobtain (9) with $W^{p+1}$ given by (22).

Given $W^{p+1}$, (9) is linear in $B$. The difference between two solutions
must then be a closed form and the general solution for the $B$ field is
$$
B_{n-p-2}=B^{part}_{n-p-2}+\text{closed $(n-p-2)$-form} \tag 26
$$

In particular, the closed form corresponding to the difference between two
solutions obtained from different choices of the surfaces $S^{n-1}_i$ (but
with the same $W^{p+1}$) may be obtained in a form similar as the one
discussed for the case $n=3$, $p=0$. As in the case of the particle in 3
dimensions, an odd permutation in the order of the surfaces $S^{n-1}_j$ in
(23) corresponds to a solution with a different orientation of $W^{p+1}$.

More general allowed world manifolds $W^{p+1}$ are obtained by the
composition of the elementary ones given by (22). The particular solution
for the $B$ field is the superposition of the solutions corresponding to each
elementary $W^{p+1}$.

In the particular case of an ($n-2$)-brane in $M^n$ the solution (23) is
written only in terms of the ($n-1$)-dimensional world manifold $W^{n-1}$ and
reduces to $B^{part}=\Theta_{W^{n-1}}$. For $n=2$ we recover the case of a
particle in a 2-manifold discussed in Ref.[12]. For n=3 we have a string
interacting with a scalar field a an antisymmetric 2-tensor in a 3-manifold.
It is also possible to see that the assumptions we have considered are also
necessary requirements. In fact, the $B$ field must be expressed in terms of
step functions in order to obtain the $\delta$ singularities on $W^{p+1}$
necessary to fulfill eq. (9).

We have thus been able to obtain a compact expression for the general solution
to the field equations of a $p$-brane sweeping a (p+1)-dimensional world
sub-manifold in $M^n$ interacting with a $BF$ topological theory..  The world
manifolds are rstricted to be the intersection of $n-(p+1)$ submanifolds
$S^{n-1}_i$ homologous to zero. For the case of particles in two dimensions
this reduces to the requirement of the world lines being homologous to zero
discussed  in Ref.[12].The solution
obtained is closely related to the existence of  step functions defined
globally on $M^n$.
It is interesting to mention that the field configurations obtained for the
$B$ fields are similar to the ones that appear in the computation of the
correlation functions of disorder variables in field theoretical models.
Disorder variables [13] when non trivial are related to soliton operators in
those models [14]. They are non-local quantities depending in the case of
models with point-like solitons on $(n-2)$-dimensional submanifolds. In the
functional approach their correlation functions are obtained introducing
$(n-1)$-dimensional defects in the action.[13]. After imposing independence of
the correlation function on the position of the defects they can be written as
the coupling of the original model with an external field of the form (23).

\vskip .5cm
\item{}{\bf References}
\item{1.-}E. Witten Commun. Math. Phys. {\bf 117} (1988) 353; ibid {\bf 121}
(1989) 351.
\item{2.-}E. Guadagnini, M. Martellini and M. Mitchev Nucl. Phys. {\bf B330}
(1990) 575; P. Cotta-Ramusino et al Nucl. Phys. {\bf B330} (1990) 557;
B. Broda Phys. Lett. {\bf B254} (1991) 111.
\item{3.-}L. F. Cugliandolo, G. Lozano and F. A. Shaposnik Phys. Lett.
{\bf 244B} (1990) 249.
\item{4.-}D. Birmingham, M. Rakowski and G. Thompson Phys. Lett. {\bf B214}
(1988) 381; {\it ibid} {\bf 212B} (1988) 187.
\item{5.-}M. Caicedo and A. Restuccia Phys.Lett {\bf B307} (1993) 77
\item{6.-}L. Baulieu and B. Grossman Phys. Lett. {\bf B212} (1988) 351.
\item{7.-}F. Wilczek Phys. Rev. Lett. {\bf 48} (1982) 1144; ibid {\bf 44}
(1982) 957; R. Mackenzie and F. Wilczek Int. Jour. Mod. Phys. {\bf A} (1988)
2827.
\item{8.-}A. M. Polyakov Mod. Phys. Lett. {\bf A3} (1988) 325.
\item{9.-}A. S. Schwartz Lett. Math. Phys. {\bf 2} (1978) 247; M. Blau and
G. Thompson Ann. Phys. (NY) {\bf 205} (1991) 130;G..Horowitz Commun Math Phys
{\bf 125} (1989) 417
\item{10.-}X. Fustero, R. Gambini and A. Trias Phys. Rev.
Lett. {\bf 62}  (1991) 1964
\item{11.-}D. Birmingham, M. Blau, M.Rakowski and G. Thompson Phys. Rep.  {\bf
209} (1991) 129.
\item{12.-}J.P.Lupi and A.Restuccia Phys. Lett. {\bf B309} (1993) 53
\item{13.-}L.Kadanoff an H.Ceva Phys. Rev. {\bf B3} (1971) 3918.
\item{14.-}E.C.Marino and J.A.Swieca Nucl. Phys. {\bf B170} [FS1] (1980)
175;E.C.Marino Phys. Rev. {\bf D38} (1988) 3194;E.C.Marino and
J.E.Stephany Ruiz  {\it ibid} {\bf D39} (1989) 3690.
\bye